\documentclass[runningheads]{llncs}
\usepackage{graphicx}
\newcommand{\cmark}{\ding{51}}%
\newcommand{\xmark}{\ding{55}}%
\usepackage{xcolor,soul}

\usepackage{amssymb}
\usepackage{pifont}

\PassOptionsToPackage{hyphens}{url}\usepackage{hyperref}
\usepackage{amsmath}
\usepackage{algorithm}
\usepackage[noend]{algpseudocode}
\usepackage{utfsym}
\usepackage{xcolor,colortbl}
\usepackage{subcaption}
\usepackage{fontawesome5}

\begin{document}
%
\title{Keep your memory dump shut: Unveiling data leaks in password managers}
%
%
\author{Efstratios Chatzoglou\inst{1}\orcidID{0000-0001-6507-5052} \and
Vyron Kampourakis \inst{2} 
\orcidID{0000-0003-4492-5104} \and
Zisis Tsiatsikas \inst{1}\orcidID{0000-0002-9481-0906} \and
Georgios Karopoulos \inst{3}\orcidID{0000-0002-0142-7503}\faIcon{envelope} \and
Georgios Kambourakis \inst{1}\orcidID{0000-0001-6348-5031}}

\authorrunning{E. Chatzoglou et al.}
%
\institute{University of the Aegean, 83200 Karlovasi, Greece \email{\{efchatzoglou, tzisis, gkamb\}@aegean.gr} \and Norwegian University of Science
and Technology, 2802 Gjøvik, Norway \email{vyron.kampourakis@ntnu.no} \and European Commission, Joint Research Centre (JRC), Ispra, Italy \email{georgios.karopoulos@ec.europa.eu}}

%
\maketitle 

\begin{abstract}
Password management has long been a persistently challenging task. This led to the introduction of password management software, which has been around for at least 25 years in various forms, including desktop and browser-based applications. This work assesses the ability of two dozen password managers, 12 desktop applications, and 12 browser-plugins, to effectively protect the confidentiality of secret credentials in six representative scenarios. Our analysis focuses on the period during which a Password Manager (PM) resides in the RAM. Despite the sensitive nature of these applications, our results show that across all scenarios, only three desktop PM applications and two browser plugins do not store plaintext passwords in the system memory. Oddly enough, at the time of writing, only two vendors recognized the exploit as a vulnerability, reserving CVE-2023-23349, while the rest chose to disregard or underrate the issue.


\end{abstract}
\section{Introduction}
\label{S:Introduction}

Twenty years after Bill Gates predicted the death of passwords~\cite{gates}, here we are still using them. What is more, there are still major data breaches that are due to weak or insecure passwords, such as in the recent cases of SolarWinds~\cite{solarwinds} and Colonial Pipeline~\cite{colonial}. Passwords come in different forms, such as personal identification numbers (PINs), passphrases, and frequently as one of the components of multifactor authentication. In between secure but impractical solutions, such as remembering all passwords, and practical but insecure ones, such as using the same password everywhere, one of the most popular ways for protecting credentials is password managers (PMs). Moreover, even when using PMs, usability clearly affects the security and trustworthiness of such applications~\cite{CHAUDHARY201969}. Thus, the main question behind our study is: 25 years after the development of the first PM~\cite{passsafe}, how secure are our passwords when stored in such applications?

Looking first at how software developers and vendors handle PM security, as software applications, PMs capitalize on relevant principles, methodologies and tools, as well as standards and best practices for secure software development. Nevertheless, absolute security cannot be guaranteed, even when using all the above means. Moreover, one of the main building blocks of a secure software development lifecycle is risk management. During this process, however, software vendors need to decide which security issues to tackle, wittingly ignoring in several cases those that they consider of minor importance or extravagant to address. In fact, during our study, out of the seven PM vendors contacted, only two recognized the issues reported in this paper and took action to remediate them.

Moving to how PM end-users perceive the value of security, the work in~\cite{Pearman19} suggests that users of standalone password managers may prioritize security, whereas users of built-in managers seem to value convenience more. The work in~\cite{Fagan2017} compared users of password managers with non-users of such applications, finding that the main motivation of password manager users was convenience, instead of security, and the impeding factor for non-users was security concerns. These findings suggest that PM users do not always put security as their first priority when adopting such applications.

There are frequent mentions of PMs in the media in the form of advertisements, comparisons or security analyses~\cite{bestmanagers1}, attempting to help the user select the most suitable PM. However, as absolute security cannot be guaranteed, it turns out that sporadically even popular PMs, which are recognized for the level of security they provide, can suffer from security issues~\cite{bitdefender}. One such example is a recent critical vulnerability affecting KeePass, which allowed the attackers to recover the master key~\cite{keepass} safeguarding the PM's vault. In this context, this paper presents a security analysis of 12 desktop and 12 browser plugin PMs to understand the level of private information they potentially leak. Our study is focused on the period during which each of these applications is loaded on the system memory.

The contributions of this work can be summarized as follows:

\begin{itemize}
    \item We examine which modern PMs allow the extraction of plaintext credentials from the RAM, offering exploits that can streamline the retrieval process.
    \item We investigate to which extent the leaking information reveals repetitive patterns that augment the attack surface.
    \item We examine PM vendor conformity to OWASP secure development guidelines with respect to the exposure of private information.
    \item Following responsible disclosure, we notified the respective PM vendors about the  identified vulnerabilities and succinctly discuss our interaction with them.
\end{itemize}

The rest of the paper is structured as follows. The next section details the threat model. 
Section~\ref{S:Experiments} presents our methodology, the derived results, and ways of exploitation in the context of red teaming. Section~\ref{S:Best:Practices} elaborates on best practices, including possible mitigation strategies. The related work is given in Section~\ref{S:Related:Work}. The last section concludes and offers directions for future work.

\section{Threat model}
\label{S:Threat:Model}

Generally, data leakage attacks directly violate the first two properties of the Confidentiality, Integrity, and Availability (CIA) triad. For PMs, however, availability is also indirectly affected in case an attacker obtains the PM's master password and subsequently changes it, locking the legitimate user outside the PM's vault. Similarly, a user can be locked outside a service account if an attacker changes the user's password. Essentially, in the case of PMs, where the stored information is considered critical, maintaining these properties is paramount. Nevertheless, as this work proves through an experimental evaluation of several renowned PMs in Section~\ref{S:Experiments}, data leakage is still feasible under certain circumstances. This section establishes the threat model under which an outsider attacker operates.

We concentrate on client-side PMs assuming that the targeted machine is situated within a local network, typically protected by a firewall, say, the one provided by default by MS Windows. In this common setting, the adversary needs to somehow penetrate the network and gain access to the victim's machine. This can be done in a plethora of ways, which however are outside the scope of the current work; for instance, the attacker might lure the victim to download and execute a piece of malware, often the result of a phishing attack.

After that, the attacker needs to dump the PM's processes; typically, these are more than one. With reference to Section~\ref{S:Experiments}, for all the 12 standalone PMs but 1Password this is straightforward; no special privileges are required. On the other hand, 1Password's processes run with a high integrity level. This means that even if the targeted user has administrative privileges, due to the User Account Control (UAC) Windows security feature, the malware (or generally the attacker's process) will often default to medium integrity, obstructing access to resources with higher integrity levels. In such a case, the attacker needs to elevate the integrity level silently, without spawning a UAC pop-up to the user. This can be accomplished through a UAC bypass method, including registry key manipulation, DLL hijacking, and elevated COM interface. Indicatively, a well-utilized DLL hijack-based method is given in~\cite{UACMe}. Last but not least, the attacker must devise a way to efficiently search each one of the acquired dump (.DMP) files, say, through some script implementing pattern matching as detailed in subsection~\ref{SS:Demonstrative:Exploit}.

\section{Experiments}
\label{S:Experiments}

The current section details our methodology and findings. Namely, after presenting the utilized testbed, we report on the different test scenarios and discuss the derived results.

\subsection{Testbed and test scenarios}
\label{SS:Testbed:And:Test:Scenarios}

Each of the examined PM applications was installed on a separate clean Virtual Machine (VM) instance running MS Windows 10; the last OS update was performed in Jan 2024. Every VM was equipped with 5GB of RAM and a quad-core CPU. For testing each PM, we compiled six distinct, representative of use scenarios, S1 to S6, as presented in Table~\ref{T:Scenarios}. Overall, these test scenarios correspond to the different states a PM application may be in relation to the actions of the end-user.

With reference to the leftmost column of Table~\ref{ST:PM:Desktop:Results}, the first round of experiments involved 12 popular PM desktop applications~\cite{bestmanagers1}. In a subsequent round, as shown in the leftmost column of Table~\ref{T:PM:Plugins:Results}, we repeated all six scenarios against 12 frequently encountered browser PM plugins~\cite{bestmanagers3}. The latter checks have been carried out on Chrome v120.0.6099.216. By comparing the above-mentioned table columns, it is obvious that half of the PM products are common in both Tables~\ref{T:Dual:Standalone:PM:Results} and~\ref{T:PM:Plugins:Results}.
All the experiments were carried out in Jan. 2024 using the latest at that time version of each PM or browser plugin.

\begin{table}[ht]
  \centering
  \resizebox{0.8\textwidth}{!}{
  \begin{tabular}{p{0.13\linewidth} | p{0.85\linewidth}}
    \hline
    \textbf{Scenario} & \textbf{Description} \\
    \hline

    S1 & Enter the master password and dump the relevant processes. \\
    \hline
    S2 & Manually lock the PM and dump the relevant processes. \\
    \hline
    S3 & After a certain amount of idle time, the PM is locked automatically; dump the relevant processes. \\
    \hline
    S4 & After creating a new entry password, dump the PM's processes. \\
    \hline
    S5 & While the PM is unlocked, click on a random entry in the corresponding list and dump the relevant processes. \\
    \hline
    S6 & Kill the relevant processes through the task manager, rerun the application without entering the master password and dump the relevant processes. \\
    \hline
  \end{tabular}}
  
  \caption{Utilized scenarios per PM application. 
  The term ``relevant processes'' in the scenarios refers to either the PM's or the browser's processes in case of a plugin.}
  \label{T:Scenarios}
\end{table}

\subsection{Results}
\label{SS:Results}

Every dump (.DMP) file created after executing each of the scenarios of Table~\ref{T:Scenarios} was examined for possibly containing secret information in cleartext. Specifically, we look for user credentials, either the PM's master password or entry passwords; usernames are also of interest, but for the needs of this subsection the concentration is on passwords. Each PM entry corresponds to the user's sign-in credentials to some service, say, Facebook, as they are contained in the PM's vault. Typically, an entry comprises a username, password, and the URL of the corresponding service. No less important, normally, a PM (or browser in the case of a plugin) spawns more than one processes, therefore a memory dump event as those in Table~\ref{T:Scenarios} may comprise numerous .DMP files. Such a file contains arbitrary information and its average size may be several megabytes, which renders its manual analysis with a hex editor cumbersome. Instead, as detailed in subsection~\ref{SS:Demonstrative:Exploit}, this can be relatively easily automated.

The results of applying all six scenarios to each standalone PM are recapitulated in Table~\ref{ST:PM:Desktop:Results}, where the~\cmark~and the~\xmark~symbols indicate either a positive or negative result regarding password leakage, respectively. Specifically, for each PM and scenario, the table shows if one or more of the respective dump files contain either or both the end-user's master and entry passwords. Note that each scenario is independent of the others, i.e., before executing each scenario, the PM application (or the browser in the case of plugins) was terminated. 

With reference to Table~\ref{ST:PM:Desktop:Results}, it is observed that only three PMs do not expose any plaintext password across all six scenarios. This does not mean that the password(s) are indeed not present in the memory (dump); contrastingly, they may exist in the dump, but they are in some secret obfuscated form that only the PM can parse. Also on the positive side, no PM leaks any plaintext password in S6, which however was somewhat expected. On the downside, four PMs expose either or both types of passwords in all scenarios but S6. In addition, three more PMs leak both types of passwords in at least three different scenarios. On top of that, half of the PMs leak the master password in S1, which is expected to be executed much more frequently than the rest of the scenarios. The latter outcome is not much better for S2 to S5, where three, four, five, and six PMs leak the master password. With reference to entry passwords, the situation is not markedly better given that in S1 to S5 at least two PMs expose entry passwords; the worst results are obtained in S4 and S5, where nine and seven PMs leak this type of password, respectively. Overall, across all the scenarios and all the PMs, 50 leaks are perceived, 24 and 26 for the master and entry passwords, respectively. Naturally, this figure cannot be seen as positive, summoning vendors up for corrective actions.

An additional important remark that emerged during our experiments with the standalone PM applications relates to the number of times a password appears in plaintext in the corresponding memory dump file. This information per scenario per PM is summarized in Table~\ref{ST:PM:Desktop:Results:occurences}. It should be clarified that for S4 and S5, the table contains the number of times the newly added (S4) or clicked on (S5) entry password is found in the respective .DMP file. From a bird's eye view, as expected, the results between the two subtables of Table~\ref{T:Dual:Standalone:PM:Results} coincide. Concentrating on Table~\ref{ST:PM:Desktop:Results:occurences}, the worst results for the entry password(s) are observed for S1, S4, and S5, where two, six, and four PMs expose at least four instances of the same password, respectively. This number even reaches double digits in S4 (for three PMs) and S5 (for one PM). Continuing from the previous result, S4 yields the worst score, demonstrating six PMs exposing at least four instances of the same newly added entry password. On the other hand, the situation regarding the master password seems better, with the poorer and best scenario in terms of security being S1 and S2, respectively. That is, in S1 three PMs expose at least four instances of the master password, while in S2 no PM exposes more than two instances. As a general comment, with special reference to S4, and to a lesser degree to S5 and S1, a significant number of PMs expose user's entry passwords too many times, which in turn increases the attacker's chances of discerning the password in the dump file, even if the latter is inspected visually.

The corresponding results referring to the 12 browser plugins are given in Table~\ref{T:PM:Plugins:Results}. As perceived, only two plugins expose zero plaintext leaks across all scenarios. Notably, however, six plugins leak a password in at least two different scenarios. Significant also is that 9 out of 12 or $\approx$75\% of the plugins leak a password in S1, which is considered the most common from a user's perspective. Altogether, across all the scenarios and all the plugins, 23 leaks are observed, which are split into 3 and 20 for the master and entry passwords, respectively. Although somewhat better vis-\`a-vis the results of Table~\ref{ST:PM:Desktop:Results}, this figure reveals again a rather unfavorable average result in terms of security engineering for this type of software.

For better appreciating the results of Table~\ref{T:PM:Plugins:Results}, the following observations are noteworthy. First, some plugins, including 1Password and Ironvest require prior interaction between the browser and themselves for leaking an entry password. Second, certain plugins, including $^\star$PM, Kaspersky, and Sticky Password mandate the installation of the homonymous desktop PM; therefore, when the desktop PM application is locked, the plugin is also locked automatically. Third, a minority of plugins, e.g., Roboform, need prior browser interaction with any saved entry URL or the user to click on the plugin's dashboard for loading all entries. Fourth, in some cases, e.g., 1Password, the leak is visible after several seconds, possibly for the browser to finish loading the respective plugin. Lastly, most plugins do not sweep PM-loaded user credentials away from the browser's process after some time has elapsed. Bitdefender and Bitwarden are the only exceptions to this situation, clearing the browser's process after 5 and 10 min respectively. Comparing the common six PMs in Tables~\ref{ST:PM:Desktop:Results} and~\ref{T:PM:Plugins:Results} it is extrapolated that, as expected due to the second point above, there are two zero-leak PMs in both their standalone and plugin version, while for the remaining ones the situation is more or less better in their plugin version.

\begin{table}[ht]
    \centering
    \resizebox{\textwidth}{!}{
    \begin{subtable}{0.55\textwidth}
        \centering
        \begin{tabular}{ | l | c | c | c c | c c | c c | c c | c c | c c |}
            \hline
    {\textbf{PM application}} & \multicolumn{2}{c|}{\textbf{S1}} & \multicolumn{2}{c|}{\textbf{S2}} & \multicolumn{2}{c|}{\textbf{S3}} & \multicolumn{2}{c|}{\textbf{S4}} & \multicolumn{2}{c|}{\textbf{S5}} & \multicolumn{2}{c|}{\textbf{S6} } \\ 
    \hline
    
    1Password & \cmark & \cellcolor[gray]{0.8}\xmark & \cmark & \cellcolor[gray]{0.8}\xmark & \cmark & \cellcolor[gray]{0.8}\xmark & \cmark & \cellcolor[gray]{0.8}\cmark & \cmark & \cellcolor[gray]{0.8}\xmark & \xmark & \cellcolor[gray]{0.8}\xmark \\ 
    \hline
    
    Bitwarden & \cmark & \cellcolor[gray]{0.8}\cmark & \xmark & \cellcolor[gray]{0.8}\xmark & \xmark & \cellcolor[gray]{0.8}\xmark & \cmark & \cellcolor[gray]{0.8}\cmark & \cmark & \cellcolor[gray]{0.8}\cmark & \xmark & \cellcolor[gray]{0.8}\xmark \\ \hline

    $^\star$PM & \xmark & \cellcolor[gray]{0.8}\xmark & \xmark & \cellcolor[gray]{0.8}\xmark & \xmark & \cellcolor[gray]{0.8}\xmark & \xmark & \cellcolor[gray]{0.8}\cmark & \xmark & \cellcolor[gray]{0.8}\cmark & \xmark & \cellcolor[gray]{0.8}\xmark\\ 
    \hline

    Kaspersky & \xmark & \cellcolor[gray]{0.8}\xmark & \xmark & \cellcolor[gray]{0.8}\xmark & \xmark & \cellcolor[gray]{0.8}\xmark & \xmark & \cellcolor[gray]{0.8}\xmark & \xmark & \cellcolor[gray]{0.8}\xmark & \xmark & \cellcolor[gray]{0.8}\xmark \\ 
    \hline
    
    KeePass 2 & \xmark & \cellcolor[gray]{0.8}\xmark & \xmark & \cellcolor[gray]{0.8}\xmark & \xmark & \cellcolor[gray]{0.8}\xmark & \xmark & \cellcolor[gray]{0.8}\xmark & \xmark & \cellcolor[gray]{0.8}\xmark & \xmark & \cellcolor[gray]{0.8}\xmark \\ 
    \hline

    KeePassXC & \xmark & \cellcolor[gray]{0.8}\xmark & \xmark & \cellcolor[gray]{0.8}\xmark & \xmark & \cellcolor[gray]{0.8}\xmark & \xmark & \cellcolor[gray]{0.8}\cmark & \xmark & \cellcolor[gray]{0.8}\xmark & \xmark & \cellcolor[gray]{0.8}\xmark \\ 
    \hline
    
    Keeper & \cmark & \cellcolor[gray]{0.8}\cmark &  \xmark & \cellcolor[gray]{0.8}\xmark & \cmark & \cellcolor[gray]{0.8}\cmark & \cmark & \cellcolor[gray]{0.8}\cmark & \cmark & \cellcolor[gray]{0.8}\cmark & \xmark & \cellcolor[gray]{0.8}\xmark \\ 
    \hline

    Nordpass & \cmark & \cellcolor[gray]{0.8}\xmark & \xmark & \cellcolor[gray]{0.8}\xmark & \xmark & \cellcolor[gray]{0.8}\xmark & \xmark & \cellcolor[gray]{0.8}\cmark & \cmark & \cellcolor[gray]{0.8}\cmark  & \xmark & \cellcolor[gray]{0.8}\xmark\\ 
    \hline

    Passwarden & \cmark & \cellcolor[gray]{0.8}\cmark & \cmark & \cellcolor[gray]{0.8}\cmark & \cmark & \cellcolor[gray]{0.8}\cmark & \cmark & \cellcolor[gray]{0.8}\cmark & \cmark & \cellcolor[gray]{0.8}\cmark  & \xmark & \cellcolor[gray]{0.8}\xmark\\ 
    \hline

    PasswordBoss & \xmark & \cellcolor[gray]{0.8}\cmark & \xmark & \cellcolor[gray]{0.8}\cmark & \xmark & \cellcolor[gray]{0.8}\cmark & \xmark & \cellcolor[gray]{0.8}\cmark & \xmark & \cellcolor[gray]{0.8}\cmark & \xmark & \cellcolor[gray]{0.8}\xmark \\ 
    \hline

    RoboForm & \cmark & \cellcolor[gray]{0.8}\cmark & \cmark & \cellcolor[gray]{0.8}\xmark & \cmark & \cellcolor[gray]{0.8}\xmark & \cmark & \cellcolor[gray]{0.8}\cmark & \cmark & \cellcolor[gray]{0.8}\cmark & \xmark & \cellcolor[gray]{0.8}\xmark \\ 
    \hline
    
    StickyPassword & \xmark & \cellcolor[gray]{0.8}\xmark & \xmark & \cellcolor[gray]{0.8}\xmark & \xmark & \cellcolor[gray]{0.8}\xmark & \xmark & \cellcolor[gray]{0.8}\xmark & \xmark & \cellcolor[gray]{0.8}\xmark & \xmark & \cellcolor[gray]{0.8}\xmark \\ 
    \hline
        \end{tabular}
        \caption{Leaks per type of password per scenario}
        \label{ST:PM:Desktop:Results}
    \end{subtable}%
    \begin{subtable}[b]{0.75\textwidth}
        \centering
        \begin{tabular}{ | l | c c | c c | c c | c c | c c | c c |}
            \hline

    {\textbf{PM application}} & \multicolumn{2}{c|}{\textbf{S1}} & \multicolumn{2}{c|}{\textbf{S2}} & \multicolumn{2}{c|}{\textbf{S3}} & \multicolumn{2}{c|}{\textbf{S4}} & \multicolumn{2}{c|}{\textbf{S5}} & \multicolumn{2}{c|}{\textbf{S6}}\\ 
 
    \hline 
    1Password & 10 & \cellcolor[gray]{0.8}- & 2 & \cellcolor[gray]{0.8}- & 2 & \cellcolor[gray]{0.8}- & 2 & \cellcolor[gray]{0.8}4 & 1 & \cellcolor[gray]{0.8}- & \multicolumn{2}{c|}{ - }\\ 
    \hline

    Bitwarden & 8 & \cellcolor[gray]{0.8}2 &  - &  \cellcolor[gray]{0.8}- & - & \cellcolor[gray]{0.8}- & 1 & \cellcolor[gray]{0.8}7 & 1 &  \cellcolor[gray]{0.8}3 & \multicolumn{2}{c|}{ - }\\ \hline

    $^\star$PM & - & \cellcolor[gray]{0.8}- &  - &  \cellcolor[gray]{0.8}- & - & \cellcolor[gray]{0.8}- & - & \cellcolor[gray]{0.8}1 & - & \cellcolor[gray]{0.8}1 & \multicolumn{2}{c|}{ - }\\ \hline

    Kaspersky & - & \cellcolor[gray]{0.8}- &  - & \cellcolor[gray]{0.8}-  & - & \cellcolor[gray]{0.8}- & - & \cellcolor[gray]{0.8}- & - & \cellcolor[gray]{0.8}- &  \multicolumn{2}{c|}{ - } \\ \hline

    KeePass 2 & - &  \cellcolor[gray]{0.8}- & - & \cellcolor[gray]{0.8}- & - & \cellcolor[gray]{0.8}- & - & \cellcolor[gray]{0.8}- & - & \cellcolor[gray]{0.8}- & \multicolumn{2}{c|}{ - }\\ \hline

    KeePassXC &  - &  \cellcolor[gray]{0.8}- & - & \cellcolor[gray]{0.8}- &  - & \cellcolor[gray]{0.8}-  & - & \cellcolor[gray]{0.8}1 &  - & \cellcolor[gray]{0.8}- & \multicolumn{2}{c|}{ - } \\ \hline

    Keeper & 4 & \cellcolor[gray]{0.8}4  & - & \cellcolor[gray]{0.8}- & 4 & \cellcolor[gray]{0.8}4 & 4 & { \cellcolor[gray]{0.8}4 }  & 4 & \cellcolor[gray]{0.8}4 & \multicolumn{2}{c|}{ - }\\ \hline

    Nordpass & 2 & \cellcolor[gray]{0.8}- & - & \cellcolor[gray]{0.8}- & - & \cellcolor[gray]{0.8}- & - & \cellcolor[gray]{0.8}24 & 4 & \cellcolor[gray]{0.8}19 & \multicolumn{2}{c|}{ - } \\ \hline

    Passwarden & 2 & \cellcolor[gray]{0.8}3 & 2 & \cellcolor[gray]{0.8}3 & 2 & \cellcolor[gray]{0.8}3 & 2 &  \cellcolor[gray]{0.8}12 &  2 &  \cellcolor[gray]{0.8}8 & \multicolumn{2}{c|}{ - } \\ \hline

    PasswordBoss & - & \cellcolor[gray]{0.8}8 & - & \cellcolor[gray]{0.8}4 & - & \cellcolor[gray]{0.8}1 & - & \cellcolor[gray]{0.8}11 &  - &  \cellcolor[gray]{0.8}7 & \multicolumn{2}{c|}{ - } \\ \hline

    RoboForm & 1 & \cellcolor[gray]{0.8}1 & 1 & \cellcolor[gray]{0.8}- & 2 & \cellcolor[gray]{0.8}- & 1 & \cellcolor[gray]{0.8}2 & 1 & \cellcolor[gray]{0.8}2 & \multicolumn{2}{c|}{ - } \\ \hline  
    StickyPassword & - & \cellcolor[gray]{0.8}- & - & \cellcolor[gray]{0.8}- & - & \cellcolor[gray]{0.8}-  & - & \cellcolor[gray]{0.8}- & - & \cellcolor[gray]{0.8}- & \multicolumn{2}{c|}{ - }\\ \hline
        \end{tabular}
        \caption{Password instances per scenario}
        \label{ST:PM:Desktop:Results:occurences}
    \end{subtable}}
    \caption{Results for standalone PMs. Master and \colorbox{lightgray}{entry} cleartext password leaks are given in white and light-gray background, respectively (\cmark: password leak, \xmark: no password leak). The vendor preceded with a star ($\star)$ requested to remain undisclosed until the vulnerability has been patched.}
    \label{T:Dual:Standalone:PM:Results}
\end{table} 

\begin{table}[ht]
  \centering
  \resizebox{0.5\linewidth}{!}{
  \begin{tabular}{ | l | c c | c c | c c | c c | c c | c c |}
    \hline
    
    {\textbf{Browser PM plugin}} & \multicolumn{2}{c|}{\textbf{S1}} & \multicolumn{2}{c|}{\textbf{S2}} & \multicolumn{2}{c|}{\textbf{S3}} & \multicolumn{2}{c|}{\textbf{S4}}& \multicolumn{2}{c|}{\textbf{S5}}& \multicolumn{2}{c|}{\textbf{S6}}\\ 
    \hline 
 
    1Password  & \xmark & \cellcolor[gray]{0.8}\cmark  & \xmark & \cellcolor[gray]{0.8}\cmark & \xmark  & \cellcolor[gray]{0.8}\cmark  & \xmark  & \cellcolor[gray]{0.8}\xmark  &  \xmark & \cellcolor[gray]{0.8}\cmark & \xmark & \cellcolor[gray]{0.8}\xmark\\ \hline

    Avira  & \xmark & \cellcolor[gray]{0.8}\cmark  & \xmark & \cellcolor[gray]{0.8}\cmark  &   \multicolumn{2}{c|}{ n/a }  & \xmark  & \cellcolor[gray]{0.8}\xmark  & \multicolumn{2}{c|}{ n/a } & \xmark & \cellcolor[gray]{0.8}\cmark \\   \hline

    Bitdefender & \cmark & \cellcolor[gray]{0.8}\xmark  & \cmark & \cellcolor[gray]{0.8}\xmark  &  \multicolumn{2}{c|}{ n/a } & \xmark  & \cellcolor[gray]{0.8}\xmark  & \xmark  & \cellcolor[gray]{0.8}\cmark & \xmark & \cellcolor[gray]{0.8}\xmark \\ \hline

    Bitwarden & \xmark & \cellcolor[gray]{0.8}\cmark  & \xmark & \cellcolor[gray]{0.8}\xmark  &  \multicolumn{2}{c|}{ n/a }  & \xmark  & \cellcolor[gray]{0.8}\xmark  &    \multicolumn{2}{c|}{ n/a } & \xmark & \cellcolor[gray]{0.8}\xmark \\ \hline
    
    Dashlane & \cmark & \cellcolor[gray]{0.8}\cmark  & \xmark & \cellcolor[gray]{0.8}\xmark  &  \multicolumn{2}{c|}{ n/a }  & \xmark  & \cellcolor[gray]{0.8}\xmark  & \multicolumn{2}{c|}{ n/a } & \xmark & \cellcolor[gray]{0.8}\xmark \\ \hline

    $^\star$PM & \xmark & \cellcolor[gray]{0.8}\xmark  & \xmark & \cellcolor[gray]{0.8}\xmark  & \xmark & \cellcolor[gray]{0.8}\xmark  & \xmark & \cellcolor[gray]{0.8}\cmark & \xmark & \cellcolor[gray]{0.8}\xmark & \xmark & \cellcolor[gray]{0.8}\xmark \\ \hline
        
    Ironvest  & \xmark & \cellcolor[gray]{0.8}\cmark  & \xmark & \cellcolor[gray]{0.8}\cmark  &   \multicolumn{2}{c|}{ n/a }  & \xmark  & \cellcolor[gray]{0.8}\xmark  & \multicolumn{2}{c|}{ n/a } & \xmark & \cellcolor[gray]{0.8}\xmark \\\hline

    Kaspersky  & \xmark & \cellcolor[gray]{0.8}\xmark  & \xmark & \cellcolor[gray]{0.8}\xmark  &  \xmark & \cellcolor[gray]{0.8}\xmark  & \multicolumn{2}{c|}{ n/a }  & \xmark & \cellcolor[gray]{0.8}\xmark & \xmark & \cellcolor[gray]{0.8}\xmark \\\hline
    
    LastPass  & \xmark & \cellcolor[gray]{0.8}\cmark  & \xmark & \cellcolor[gray]{0.8}\cmark  &   \multicolumn{2}{c|}{ n/a }  & \xmark  & \cellcolor[gray]{0.8}\xmark  &  \multicolumn{2}{c|}{ n/a } & \xmark & \cellcolor[gray]{0.8}\cmark \\ \hline
    
    Norton  & \xmark & \cellcolor[gray]{0.8}\cmark  & \xmark & \cellcolor[gray]{0.8}\cmark  &   \multicolumn{2}{c|}{ n/a }  & \xmark  & \cellcolor[gray]{0.8}\xmark  & \multicolumn{2}{c|}{ n/a } & \xmark & \cellcolor[gray]{0.8}\cmark \\  \hline
    
    RoboForm   & \xmark & \cellcolor[gray]{0.8}\cmark  & \xmark & \cellcolor[gray]{0.8}\xmark  &   \multicolumn{2}{c|}{ n/a }  & \xmark  & \cellcolor[gray]{0.8}\xmark  &  \multicolumn{2}{c|}{ n/a } & \xmark & \cellcolor[gray]{0.8}\xmark   \\ \hline

    StickyPassword  & \xmark & \cellcolor[gray]{0.8}\xmark  & \xmark & \cellcolor[gray]{0.8}\xmark  &  \xmark & \cellcolor[gray]{0.8}\xmark  &  \xmark & \cellcolor[gray]{0.8}\xmark  & \xmark & \cellcolor[gray]{0.8}\xmark & \xmark & \cellcolor[gray]{0.8}\xmark \\\hline

  \end{tabular}}
  \caption{Results for PM browser plugins. Master and \colorbox{lightgray}{entry} cleartext password leaks are given in white and light-gray background, respectively (\cmark: password leak, \xmark: no password leak). n/a (not applicable) means that either the plugin does not have an automatic lock feature (S3), or the plugin does not offer an ``add new entry'' feature (S4), or the plugin loads all entries automatically when the browser is started (S5).}
  \label{T:PM:Plugins:Results}
\end{table} 

\subsection{Discussion}
\label{SS:Discussion}

Abiding by a Coordinated Vulnerability Disclosure (CVD) process, the results have been promptly communicated to most of the affected PM vendors along with the corresponding exploit(s). Two vendors did acknowledge the issue and reserved a Common Vulnerabilities and Exposures (CVE) ID, namely CVE-2023-23349, pledging to also provide a respective patch promptly. The rest of them provided arguments in two directions. The first implies that the vendors are already aware of the problem, but they largely tend to underrate it. Namely, their basic claim is that performing such an attack would require an already compromised system and a malicious actor with escalated privileges. On the one hand, this assertion is half true; based on our results in subsection~\ref{SS:Results}, there do exist PM products that do not leak secret information in plaintext, and this stands true for both the standalone PMs (three instances) and the respective plugins (two instances). On the other hand, as detailed in Section~\ref{S:Threat:Model}, all PMs except one do not even require elevated privileges for dumping their processes. In this respect, the attacker does not even need to bypass UAC, which generally is cumbersome.

In relation to the previous argument, others may reason that a keylogger malware may achieve the same goal, without collecting and inspecting a PM memory dump file. The counterargument here is that a keylogger needs to stealthily operate round-the-clock and opportunistically wait for the victim to type a password. Even more, the attacker can save the relevant .DMP file and analyze it offline, let alone that if the respective PM divulges the master password, then the attacker instantly gains access to every entry the victim keeps in the PM's vault.

The second line of arguments, even not so common, suggests that some of the PM vendors are unsure about how to address the issue effectively. That is, if they apply some form of obfuscation to the passwords loaded in memory, they may provide a false sense of security; this is a form of security by obscurity, which generally, as well as in this case, is only effective as long as the (secret) obfuscation pattern remains confidential. On the flip side, unauthorized access to the PM will typically lead to multiple breaches (this equals the number of different entry passwords stored in the PM). In the presence of no other viable and stronger defense, obfuscation seems, at least for the moment, the only way forward.  To enhance the robustness of the obfuscation procedure, a variable, frequently changing pattern should be used, ideally paired with other techniques, e.g., the system-protected process security model, also known as Protected Process Light (PPL) technology~\cite{System-protected-process}. For more information on this issue, the reader is referred to Section~\ref{S:Best:Practices}.

\subsection{Red teaming and exploits}
\label{SS:Demonstrative:Exploit}

To propel research and advance red teaming methodologies and responsible disclosure in this field, we developed an open-source tool called \textit{Pandora}, which is publicly available in a GitHub repository~\cite{pandora}. Pandora assists in harvesting user credentials, both usernames and passwords, possibly present in cleartext in the processes spawned by several well-known PMs. The tool supports MS Windows 10 standalone PMs and browser plugins; at the time of writing, it supports 14 PMs, with 18 different implementations, e.g., it can dump end-user credentials either from the desktop application or the browser plugin of the same product. The tool requires the PM to be running and unlocked. For additional and up-to-date information on the tool, the interested reader is referred to~\cite{pandora}.

Moreover, for reasons of completeness, Algorithm~\ref{A:getCreds} offers a generic exploit in pseudocode. As seen in line 3, after opening the respective .DMP file, the algorithm applies a search pattern. The latter is based on the common technique exploited by anti-malware services; a file is flagged as possibly malicious when a sequence of hex bytes, i.e., a signature, is present within the file. In this work, we exploit the same logic to identify the patterns that hold the user's credentials in the corresponding PM's process memory dump. To identify a possible pattern, first, one needs to configure each PM with different credentials, namely, master and entry passwords along with their matching usernames, if required by the PM depending on the case. Then, using a hex editor, each dump file is manually investigated for containing one or more of these entries. If some entry is found, that entry's preceding and succeeding bytes are kept as possible search patterns. Then, the investigator checks if the same bytes (pattern) stay identical across different PM states with the same or other entries. If yes, then this pattern can be used to produce an exploit for the specific PM. In Algorithm~\ref{A:getCreds}, given the pattern of line 3, lines 5-21 gather the next or previous \textit{n} bytes, and convert them to UTF-8 for revealing the password (and username) in plaintext. The number of bytes \textit{n} is defined empirically, and it may be different per PM; for Algorithm~\ref{A:getCreds} this parameter is set to 300 bytes in line 12.

\begin{algorithm}[ht]
\label{L:Algo}
\caption{getCreds}\label{A:getCreds}
\small
\begin{algorithmic}[1]
    \Function{getCreds}{}
        \State Open file ``app.dmp'' in binary mode

        \State $searchPattern \gets$ [0x80, 0x00, 0x04 ... 0x00]
        \State $foundData \gets$ empty list

        \While{not end of file}
            \State Read a byte from file and store it in variable $c$

            \If{$c =$ next element in $searchPattern$}
                \State Add $c$ to $foundData$

                \If{length of $foundData =$ length of $searchPattern$}
                    \State $extractedData \gets$ empty list
                    \State $dataCount \gets 0$

                    \While{$dataCount < 300$ and not end of file}
                        \State Read a byte from file and add it to $extractedData$
                        \State Increment $dataCount$
                    \EndWhile

                    \State $utf8ExtractedData \gets$ convert $extractedData$ to UTF-8 string

                    \State Print ``Pattern Data:'' concatenated with $utf8ExtractedData$

                    \State Save into file using \texttt{saveFile} function with parameter \\\hspace{1.85cm} $utf8ExtractedData$

                    \State Clear $foundData$
                \EndIf
            \Else
                \State Clear $foundData$
            \EndIf
        \EndWhile

        \State Close file
    \EndFunction
\end{algorithmic}
\end{algorithm}

\section{Best practices}
\label{S:Best:Practices}

The current section offers a brief description of the best practices that should be employed for sensitive information management by a PM application. As discussed in Section~\ref{S:Experiments}, the success of the attack hinges on the fact that essential information is written in an unprotected (plaintext) format by the PM application. Nevertheless, there exist numerous guidelines~\cite{10.1007/978-3-319-60588-3_3,owasp-product,owasp-crypto} that dictate the proper way for handling private information by such applications.

Primarily, as delineated by the Open Worldwide Application Security Project (OWASP), the relevant guidelines concentrate on the proper employment of cryptographic primitives~\cite{owasp-product,owasp-crypto}. In this respect, encryption and hashing algorithms are considered the mainstream tools in the quiver of the PM developers to protect private information, even though key management remains an important matter in the whole process. Along the same line, obfuscation techniques can be used to hide essential information when loaded in memory; note that the obfuscation pattern must remain secret, which indeed is a form of security by obscurity. Moreover, the pattern should be frequently changed, which also entails management operations, and naturally, there is no silver bullet that can ensure key management security horizontally. No less important, OWASP suggests that private data should be stored only when necessary during the application life cycle~\cite{owasp-crypto} and remain the minimum possible time in the RAM~\cite{owasp}. This can be achieved through spawning garbage collection requests soon after every critical operation, or by overwriting the content of a critical object once the object is no longer needed.

Another commonly used technique when it comes to the application development stage is to eliminate variable copies~\cite{10.1007/978-3-319-60588-3_3}. Precisely, modern programming languages afford several operations on variables, which essentially create copies of the same information in the memory stack; each variable assignment creates a copy of the same information. This might be the case with the scrutinized PMs, as Table~\ref{ST:PM:Desktop:Results:occurences} confirms that both the master and entry passwords were identified multiple times in the respective memory dumps. In turn, this facilitates an attacker to monitor the repeating piece of data, subsequently revealing repeating private information and the associated byte patterns. Another relevant point is that some PMs, including Keeper do provide an option for wiping secret data after the PM is automatically locked. Nevertheless, this option is provided as opt-in and it is disabled by default, which does not abide by the well-known principle of establishing security defaults, namely establish default secure settings and possibly provide opt-in and opt-out. Even more, a minority of PMs support Two-Factor Authentication (2FA), which is positively appraised. Nevertheless, in some of them, the 2FA token created during user sign-in with the PM leaks in the memory as well.

Last but not least, as already pointed out in Sections~\ref{S:Threat:Model} and~\ref{SS:Discussion}, developers should also take advantage of the UAC mandatory access control enforcement feature, which limits the ability of malicious code to execute with administrator privileges. By means of UAC, every application that needs the administrator access token must prompt the end-user for approval, the so-called ``elevation prompt''. Typically, when a user runs a process, that process executes with their token and can do anything the user can do; in short, a token says who you are and what you are allowed to do. In this context, every time a Windows thread wishes to access a protected object, the operating system conducts a security check. With reference to subsection~\ref{SS:Discussion}, vendors can potentially capitalize on the so-called Protected anti-malware services, also known as PPL~\cite{System-protected-process}. This enables specially-signed programs to run in a way that they are invulnerable to tampering and termination, even by administrative users. On the downside, running a PM with administrative rights to enable PPL, may open the door for privilege escalation attacks through DLL sideloading, e.g., as given in CVE-2023-48861.

\section{Related Work}
\label{S:Related:Work}

This section elaborates on the related work on PM security, concentrating on publications from the past decade. The first set of related works focuses on applications that expose sensitive data while running on different operating systems, namely Windows, Linux and Android.
\cite{apostolopoulos2013discovering} investigated the potential leakage of private data from Android applications residing in RAM. According to the findings, 29 out of 30 applications were found to disclose information that facilitated the retrieval of user passwords. A similar research in~\cite{Lee2019TotalRP} presented an analysis of Android applications, including PMs, regarding the persistence of passwords in system memory even after they are not needed. Based on their results, one of the main components responsible for leaking private information is UI widgets. To address this, they proposed an enhancement to the Android platform by introducing the SecureTextView, a secure version of Android TextView, which employs the zeroization of specific buffers. In a related research~\cite{7784634}, the authors explored zeroization techniques across MS Windows and Linux environments. They concluded that only the MS Windows operating system provides a reliable function that can be used for this purpose. Finally, they showcased credential extraction from memory allocated to browsers, highlighting potential security vulnerabilities.

Extending the analysis to web-based PMs that offer standalone applications and/or browser extensions, the authors of~\cite{oesch2020then} conducted a security evaluation of seven web-based PMs. Their evaluation covered both the standalone application and browser extension-based versions, encompassing the entire PM lifecycle, from password generation and storage to autofill functions. Despite the improvement of the scrutinized PMs in terms of security compared to prior evaluations, persistent issues remain. These include unencrypted metadata, insecure defaults, and vulnerabilities to clickjacking attacks. The authors explored the methods that each of the PMs utilizes to securely store client data in the respective database files. However, they overlooked the manner in which the very same data are treated by the host's OS and how they are stored in RAM.

In the same vein, the authors in~\cite{li2014emperor} performed a security analysis of five web-based PMs, uncovering varying susceptibilities across all the examined PMs. Among the identified vulnerabilities were issues related to logic and authorization, misinterpretation of the web security model, and sensitivity to CSRF/XSS attacks. Nevertheless, the authors specifically focused on vulnerabilities stemming from the web-based nature of PMs. In contrast, our study takes a different angle by delving into the local security posture of PMs, with a particular emphasis on examining how sensitive data are stored within the users' RAM.
Another study~\cite{Barten2019ClientsideAO} presented flaws in the LastPass browser extension. In this work, the authors performed a password extraction attack directly from the extension's memory while the extension was active. In their research, Zhao, Yue, and Sun~\cite{Zhao2013} provided a vulnerability analysis of LastPass and Roboform, which both provide browser extensions for operating the software and cloud storage of the passwords. They identified potential threats, including the storage of credentials in plaintext locally and/or on cloud servers. Additionally, they provided recommendations to enhance data protection measures, improving the overall security level of these products.

A second set of works focused on data leakage during the autofill function that modern PMs offer. The authors of~\cite{Silver14} studied how different PMs handle automatic filling of passwords. Their work encompassed three categories of password managers: (a) built into browsers, (b) used in mobile devices, and (c) 3rd party applications. Their findings suggested that the various autofill policies followed by these tools can expose users to serious risks, such as a remote attacker stealing multiple passwords from their PM without their knowledge. Similarly,~\cite{gangwal2023autospill} presented a technique, called AutoSpill, to leak users' saved credentials during an autofill operation on a webpage loaded into an application's WebView.

As a third tendency, some authors have followed forensic techniques to study the security of PM applications. In~\cite{Gray16}, three desktop PMs were analyzed, namely KeePass, Password Safe, and RoboForm, following a systematic forensic analysis approach. The findings of this study showed that unencrypted passwords and sensitive data were found in the clipboard, Temp folders, Page files, or the Recycle Bin. Moreover, the authors of~\cite{Sabev22} followed a digital forensic procedure to analyze the storage security of two Android PM applications. Their focus is, however, on protecting long-term data stored in Android devices, leaving out attacks targeting encryption keys in main system memory.

In a similar direction to our work,~\cite{Luevanos17} scrutinized the security of three open-source PMs against various types of attacks, both new and already known. The PMs they studied were Passbolt, Padlock, and Encryptr; among other vulnerabilities, the authors found that all three applications were susceptible to keyloggers and clipboard attacks. Additionally, the authors of~\cite{carr20} analyzed five popular PMs against already disclosed and newly discovered vulnerabilities. Notably, their findings indicated that four of the five studied PMs, namely Dashlane, LastPass, Keeper, and RoboForm, were prone to clipboard attacks.

When it comes to tools for password recovery from PMs, LaZagne~\cite{lazagne} is an open-source application capable of retrieving locally stored computer passwords. It supports various methods that applications use to store their passwords, such as plaintext, APIs, custom algorithms, and databases. Among the supported applications are popular browsers, mail clients, chat apps, databases, and memory dumps from password managers such as Keepass. \textit{Pandora}~\cite{pandora}, the tool we have developed for our experiments, centers specifically on PMs and delves deeper than LaZagne. It uncovers passwords that are inadequately managed by 18 different implementations of PMs, spanning Windows apps, browsers, and browser plugins.

\section{Conclusions}
\label{S:Conclusions}

Password credentials are still the dominant way to perform user authentication. Nevertheless, with the numerous accounts that the average user manages today, it is challenging to create secure passwords while at the same time remembering them by heart. PMs come with the pledge to fill this gap by proposing strong passwords and storing them securely on desktop and mobile devices. But leaving aside the convenience provided by such applications and the widely held opinion that they are safe to use, can they really be trusted to fully deliver what they promise?

This paper provides an up-to-date security analysis of the main desktop and browser plugin PMs. This is done under the strong but viable assumption that the adversary has somehow gained access to the local system, being also capable of dumping the respective processes. Our findings suggest that, across six representative from an end-user's viewpoint scenarios, 75\% of desktop PMs and 83\% of browser PM plugins store passwords in plaintext in the system memory while in use. Among others, the key issue with these security-sensitive applications is that this ``storing'' is done in a way that creates patterns, which can be easily discerned by the adversary. Even worse, this secret information not only remains in the memory for a significant time after serving its purpose, but it is also stored in multiple copies, making the attacker's job easier. A possible direction for future work is to expand this research to mobile platforms and to several other popular applications of different kinds, including financial/banking, team collaboration, smart home, etc.

\section*{Acknowledgements}

This work is supported by the Research Council of Norway through the SFI Norwegian Centre for Cybersecurity in Critical Sectors (NORCICS) project no. 310105 and by the European Union through the Horizon 2020 project PERSEUS (Grant No. 101034240).

%
%
\bibliographystyle{splncs04}
\bibliography{references}

\end{document}